\documentclass[12pt]{article}
\usepackage{epsfig}
\begin{document}
\pagestyle{empty}
\begin{flushright}
{CERN-TH/99-389}
\end{flushright}
\vspace*{5mm}
\begin{center}
{\bf VECTOR FIELDS, FLOWS AND LIE GROUPS OF DIFFEOMORPHISMS}
\vspace*{1cm}
 
{\bf A. Peterman 
}\\
\vspace{0.3cm}
Theoretical Physics Division,
CERN \\
CH -- 1211 Geneva 23\\
\vspace*{2cm}
{\bf ABSTRACT}\\ \end{center}
\noindent
{The freedom in choosing finite renormalizations in quantum field theories (QFT) is
characterized by a set of parameters $\{ c_i \}, i = 1 \ldots , n \ldots$, which specify
the renormalization prescriptions used for the calculation of physical quantities. For
the sake of simplicity, the case of a single $c$ is selected and chosen mass-independent
if masslessness is not realized, this with the aim of  expressing the effect of an
infinitesimal change   in $c$ on the computed observables. This change is found to be
expressible in terms of an equation involving a vector field $V$ on the action's space $M$
(coordinates x). This equation is often referred to as ``evolution equation'' in physics.
This vector field generates a one-parameter (here $c$) group of diffeomorphisms on $M$.
Its flow $\sigma_c (x)$ can indeed be shown to satisfy the functional equation
$$ \sigma_{c+t} (x) = \sigma_c (\sigma_t (x)) \equiv \sigma_c \circ \sigma_t $$
$$\sigma_0 (x) = x,$$
so that the very appearance of $V$ in the evolution equation implies at once the
Gell-Mann-Low functional equation. The latter appears therefore as a trivial consequence
of the existence of a vector field on the action's space of renormalized QFT.}

\vspace*{4cm}
\begin{flushleft}
{CERN-TH/99-389}\\
December 1999
\end{flushleft}
\vfill\eject

\newpage
\setcounter{page}{1}
\pagestyle{plain}

\begin{flushright}
{\it To the memory of G. de Rham,\\
 my teacher  in mathematics.}
\end{flushright}

The so-called ``Renormalization Group'' (RG) in physical science was, from its early
beginnings \cite{peterman1}, the theory that describes the geometry of action space.
In this space the covariance of physical quantities turns out to be manifest.

This paper is intended to be an overview of the RG as it is used by physicists,
especially in Quantum Field Theory (QFT). However, the emphasis will be put on the
geometry (as said before) of the space of actions, with a view, among other things,
on how a functional equation, derived in \cite{peterman2}, becomes a simple and trivial consequence
of the geometrical set-up developed below by means of QFT
manipulations.

Of paramount importance has been the discovery \cite{peterman1} that renormalized
physical quantities satisfy equations in which the basic geometrical object is a
vector field defined in the action space $M$. Let $V$ be this vector field. Then the
theory of differentiable manifolds  \cite{peterman3} implies, due to the very existence of a vector
field, a set of theorems, lemmas and corollaries which exhausts all that can be said about the
renormalization group in physical applications. In general, one deals with a set of parameters $\{
c_i\}$ but, in the following, we shall restrict this set to a single parameter, denoted $t$ for
practical reasons. This restriction is mainly dictated by the fact that one wants to be able to make
direct comparisons with Ref.~\cite{peterman2} which uses a single parameter in its fixing of
renormalization prescriptions. Thus, in the single parameter case, we have amongst others the
following theorems.

\begin{itemize}
\item[]{\bf Theorem I.}

 A smooth vector field $V$ on a compact manifold $M$ generates a one-parameter group
of diffeomorphisms of  $M$.

\item[]{\bf Theorem II.}

 Suppose $V$ is a $C^r$ vector field on the manifold $M$, then for every $x \in M$,
there exists an integral curve of $V$, $t\rightarrow \sigma (t , x)$ such that 
\begin{enumerate}
\item  $\sigma (t , x)$ is defined for $t$  belonging to an interval $I (x) cR$, containing $t=0$
and is of class $C^{r+1}$ there.
\item $\sigma (0, x) = x$ for every $x \in M$
\item  Given $x \in M$, there is no $C^1$ integral curve of $V$ defined on an interval properly
containing $I (x)$, and passing through $x$ (i.e. such that $\sigma (0, x) = x$).
\end{enumerate}

From the uniqueness property 3, follows at once

\item[]{\bf Theorem III.}

 If $s, t$ and $s + t \in I (x)$, then we have the functional equation
\begin{equation}
\sigma (s  + t , x) = \sigma (t , \sigma (s, x))\;.
\label{peterEQ1}
\end{equation}
\end{itemize}

{\bf Flow:} The set of pairs $(x, t), x \in M , t\in I (x)$ is an open subset of $M\times R$
containing $x$, hence a smooth manifold $\sum$, of dimension $n +1$. The mapping $\sigma$: $\sum_v
\rightarrow X $ by $(x ,t ) \rightarrow \sigma (t ,x)$ is called the flow of the $C^1$ vector field
$V$.  If $M$ and $V$ are $C^\infty$, the flow is also of class $C^\infty$. Writing $\sigma (t , \cdot )
\equiv \sigma_t : x \rightarrow \sigma (t ,x)$; (Eq.~\ref{peterEQ1}) can be written
\begin{equation}
\sigma_{s+t} = \sigma_t \circ \sigma_s \;\;\; ; \;\;\; \sigma_0 =e \;.
\label{peterEQ2}
\end{equation}
The set of mappings $\{ \sigma_t\}$ is the one-parameter group mentioned by Theorem~I, provided, as
 is the case for our concern, $I = R$. (It is obvious that each element $\sigma_t$ has an inverse
$\sigma_{-t}$ so that $\sigma_t \circ \sigma_{-t} = e$)\footnote{For a sample of textbook'
references involving the fundamentals of differential geometry as well as rigorous proofs of the
topics advocated here and beyond these topics, see the item: \underline{Textbooks} in the references
at the end of the paper.}.

According to the few theorems given up to now, one sees that the vector field $V$ is the
\underline{key} concept, generating a flow $\sigma (t, x) \equiv \sigma_t (x)$ in $M$. Since the
$\sigma_t$, with fixed $t$, is a diffeomorphism $M\rightarrow M$, it represents the one-parameter
Abelian group according to (Eq.~\ref{peterEQ2}). Therefore, $V$ can be seen as the infinitesimal
generator of the flow group $\sigma_t$. Indeed, for $t$ infinitesimal, say $0 +\delta t$, we have the 
infinitesimal flow
$$ \sigma_{0 + \delta t} (x) = \sigma_0 (x) + \delta t V(x) + O (\delta t^2)$$
(with  $\sigma_0 (x) = x$), by Taylor expending $\sigma (\delta t ,x )$ around $\delta t = 0$, so that
$$ \frac{\sigma (0 + \delta t , x) - \sigma (0, x)}{\delta t} = V (x)\;.$$

Or, taking the limit   $\delta t \rightarrow 0$
\begin{equation}
\left.{\frac{d\sigma (t , x)}{dt}}\right|_{t=0} = V (x)\;.
\label{peterEQ3}
\end{equation}
$V(x)$ is therefore, as we said at length before, the infinitesimal generator of the flow group. Then,
by exponentiation, one gets
$$\sigma (t, x) = \exp [t V] \cdot x$$
which fulfils, as expected
\begin{itemize}
\item[a)] $\sigma (0, x) = x$
\item[b)] $\frac{d}{dt} \sigma (t, x) = V \exp [ t V] \cdot x = V (\sigma (t ,x))$
\item[c)] $\sigma (s + t ,x) = \sigma (t , \sigma (s, x))$
\end{itemize}
(Remember the operator nature of $V$, which can be abbreviated $V = V^\alpha  \frac{\partial}{\partial
x^\alpha}$.)

The elementary considerations made up to now, would be sufficient for the current applications of the
RG to QFT since, as   is well-known, the one-parameter $t$ is the logarithm of a scale $\mu$, and
$\frac{d}{dt} \rightarrow \mu \frac{d}{d\mu}$, which is the Abelian generator of the one-parameter
group for scale transformations of the subtraction point.

The appearance of the arbitrary dimensional parameter $\mu$\footnote{In physics, $\mu$ is generally
called ``subtraction point''.} cannot be avoided and is the source of the breakdown of conformal and
scale invariances in classical conformal invariant Lagrangians. The deep source of this anomalous
breakdown is traced back, as  is well-known, in the procedure of the second quantization
\cite{peterman4}. In this reference, R.~Jackiw notices `` $\ldots$ we may say that our present point
of view towards scale and conformal symmetry breaking was prefigured by Bohr's intuition concerning
effects of quantization on space-time symmetries.''

In conclusion, in order to fix the ideas in the present ``physical'' notations, $V$ is expressed as
$$V = \beta^a (g) \cdot \frac{\partial}{\partial g^a}\;\;\;;\;\;\; a =1 \ldots \kappa\;,$$
if there are $\kappa$ couplings in the considered theory. $g \equiv \{ g^a\}$ is the set of couplings
and $\{\beta^a(g)\}$ that of the components of the vector field.

We see therefore that the $\beta^a$ are the components $V^\alpha$ of $V$ and the $g^a$, the
coordinates $x^\alpha$ of $M$ (called in physics the action space). Finally, the components of the
flow in this space are denoted $\bar g^a (t, g)$, corresponding to the $\sigma^\alpha (t,x)$
components of the flow $\sigma (t ,x)$ on $M$.

So that, since, as we have seen, $V$ is the infinitesimal generator of the one-parameter flow group on
$M$, $\beta^a \frac{\partial}{\partial g^a}$ is the one-parameter flow group infinitesimal generator
on action space. As a flow, $\bar g(t ,g)$ can be obtained from the exponentiation of $V= \beta^a
\frac{\partial}{\partial g^a}$ (see the example in the Appendix~A). It therefore satisfies the
functional equation~(\ref{peterEQ1}) as expected, namely:
$$\bar g (t+s, g) = \bar g (t, \bar g (s , g))$$
with $\bar g (0, g) = g$ as boundary condition.\footnote{In Ref.~\cite{peterman2}, $\bar g (t, g)$ is
expressed as  $e^2 d (t , e^2)$, with $t=\log (\kappa^2/\lambda^2) \equiv \log x^2$ in their notations.
$e^2$ stands for $g$ up to a change of coordinate  since the QED case is investigated, with one
\underline{single} parameter
$\lambda$, as in the formulation of the present paper.}

The fundamental equation for QED $S$-matrix elements, already quoted in its simplest formulation in
the Abstract of Ref.~\cite{peterman1}, and which has been mentioned in this paper as the equation
introducing in physics a vector field \underline{V} was
\begin{equation}
\left.{\frac{\partial}{\partial c_i} S (x \ldots m , e, c_i) }\right|_{c_i = 0} 
 = h_{i.e.} (e) \frac{\partial}{\partial e} S(x \ldots ,
m , e)
\label{peterEQ4}
\end{equation}
or, in the simplified case (one single $c$: $c_0 = t$) of this paper
\begin{equation}
\left.{\frac{\partial}{\partial t} S (p_j \ldots m , e, t)}\right|_{t=0} = \underline{V} \cdot
S(p_j \ldots m , e)
\end{equation}
(the $p_j$ being a conjugate momenta of the $x^\prime_s$ in (Eq.~\ref{peterEQ4})), with of course
$\underline{V} = h_0 (e) \frac{\partial}{\partial e}$.

Notice that the index $i$ has been dropped in the formulation of this paper, since it stands for the
numbering  of the various arbitrary normalization conditions. 

To our knowledge, the first author who took into consideration the general case \cite{peterman1} with
several parameters $c_i$ is Crewther \cite{peterman5}. His analysis is confined to a finite set of
normalization conditions $R(c_i)$, and he put  forward the very simple argument that  it is
sufficient to consider transformations in the $c_i$-space possessing the group property, which
warrant  a satisfactory rule for this special subset $R(c_i)$ of normalization conditions to have the
group property.  This is what was called ``normalization group'' in Ref.~\cite{peterman1}. So the
group property is a feature of many subsets  of the whole set of prescriptions, $G$, but certainly
$G$ itself (the countable infinite set of prescriptions) does not, strictly speaking, possess this
property.

A very popular set of prescriptions are the so-called ``mass independent schemes'', for which the
normalization factors are computed with the bare mass set  equal to zero. Then the renormalized mass
is treated like a renormalized coupling \cite{peterman6}. In accordance with na{\"\i}ve dimensional
analysis, the vector field components can only depend on the couplings \cite{peterman7}. In this set
one finds, among others, dimensional regularization supplemented by minimal subtraction (MS) or its
cousin $\rm \overline{MS}$. This mass-independent set of prescriptions
possesses the group property. Several authors
\cite{peterman8}  tried to tackle the case when the $c_i$ are infinite in number, especially with the
aim of optimizing the perturbation series truncated at a given order. Although the possible group
property in these extreme cases has not been addressed, these authors established that the most
general coupling constant $g (c_1 , c_2, \ldots c_\infty)$ depends, as we wrote, on the countable
infinity of $c_i$ and were able to show in a particular case that the $c_i$ are linear in the $b_i$,
the numerical coefficients in the expansion in $g$ of the vector field component $\beta (g) =
\sum^\infty_{i=n} b_i g^i, n=1,2\ldots$. These coefficients are well-known for their dependence on the prescription used. In
other words, geometrically, they depend on the choice of the coordinate $\{g ^a\}$ in action space.
However, the choice of a system of reference is arbitrary and the above results do not shed light on
 which sets, if any, enjoy the group property in a Banach space.

In conclusion, the passage from a single parameter $c_0 \equiv t$ considered in this paper, to several
$c_i$, or an infinity of them, is \underline{not} straightforward. Again we might be able to
consider, as was done in \cite{peterman1}, sets of transformations in the $c_i$ space which possess
the group property. This problem involves the theory of several parameter Lie groups of
transformations and lies beyond the modest scope of this paper. Nevertheless, a few guidelines will be
given in Appendix~B in a very concise and not mathematically rigorous way.

\newpage
\setcounter{section}{0}
\setcounter{equation}{0}
\renewcommand{\thesection}{Appendix \Alph{section}}
\renewcommand{\theequation}{A.\arabic{equation}}
\section{}

As a very simple example, we take, for the vector field $V\Longleftrightarrow\beta (g)
\frac{\partial}{\partial g}$ in a one-dimensional action space (coordinate $g$), the first term of
$\beta (g)$ in a $g$ expansion, say
$$\beta (g) = b g^2\;.$$
The exponential $\exp \{ t \underline{V}\} \cdot x$ is defined by its Taylor expansion
$$\exp \{ t \underline{V}\} = \sum^\infty_{n=0} (t\underline{V})^n \frac{1}{n!}$$
From 
$$\underline{V}\,\underline{V}\rightarrow bg^2 \frac{\partial}{\partial g} \left({bg^2
\frac{\partial}{\partial g}}\right) = 2b^2 g^3 \frac{\partial}{\partial g} + b^2 g^4
\frac{\partial^2}{\partial g^2}$$ 
it is straightforward to deduce
$$\underbrace{\underline{V }\,\underline{V} \ldots \underline{V}}_{n \,\mbox{factors}} = n! b^n g^{n+1}
\frac{\partial}{\partial g} + O \left({\frac{\partial^n}{\partial g^n} \;,\; n \geq 2}\right)\;.$$
Therefore
$$\bar g (t,g) = \exp \{ t \underline{V}\} g = \sum^\infty_{n=0} t^n b^n g^{n+1} = \frac{g}{1-tbg}\;,$$
a well-known result.

As an exercise, the reader can establish, according to the above, the approximate Bogoljubov-Shirkov
relation  in the next order  for $\bar g (t , g)$ \cite{peterman9}
$$\bar g (t, g) = g[1-b_1 gt + \frac{b_2}{b_1} g \log (1-b_1 gt)]^{-1}$$
by taking
$$\beta (g) = b_1 g^2 + b_2 g^3\;.$$

A second exercise is to show that
$$\exp \{ t \underline{V}\} g^n = \bar g^n (t,g) = \frac{g^n}{(1-gbt)^n}$$
when the vector field $\underline{V} = \beta (g) \frac{\partial}{\partial g}$ is approximated by the
first term in the $g$ expansion of $\beta (g)$ i.e. $\beta (g) = b g^2$, like in the first example.

Since $S$-matrix elements can be expanded in powers of $g$
$$S(p_i, g)= \sum^\infty_{n = 0} a_n (p_i) g^n$$
it follows that
$$\exp \{ t \underline{V} \} \cdot S (\ldots g) = S(\ldots \bar g)$$
($p_i$ and the dots stand for arguments other than $g$ and independent of it. The case when other
couplings, like $g_i$ and masses $m_i$ occur, goes outside the one-dimensional action space and
$\underline{V}$ becomes $V= V^\alpha \frac{\partial}{\partial x^\alpha}$, the $x^\alpha$ being the
coordinates in the enlarged action space, namely the $g_i$ and $m_i$ above.)

\section{}
The passage from a one-parameter case to the case with several parameters $c_i$ is far from trivial,
although treated at length in the Textbook references, especially
[T.1]--[T.4]. Sketching what happens, from an element $g$ depending on one
parameter
$g(t) \cdot g(s) = g (s+t)$ with $g(t) = 1 + t\underline{V}$ for $t$ infinitesimal one goes to
$$ g (t_1, \ldots t_i) = 1 + t_1 \underline{V}_1 + t_2 \underline{V}_2 + \ldots t_i \underline{V}_i$$
with all $t_i$ infinitesimal. ($i$ generators $\underline{V}_i$)

The combination of two such elements $g (t_1, \ldots t_1)\cdot g (s_1, \ldots
s_i)$ is given by the well-known Baker-Campbell-Hausdorf formula.

For the product to be also an element of the set, the condition $[\underline{V}_i, \underline{V}_j] =
c_{ij}^\kappa \underline{V}_\kappa$ is necessary and sufficient. It was a condition explicitly
formulated in \cite{peterman1} for a set of normalization conditions to form a group. For our concern,
the $g (t)$ are connected with the flow $\sigma_t (\cdot)$. Therefore the group property concerns the
flows, as in the one-parameter case discussed in this paper.

The non-trivial aspect now is that we must distinguish between the left combination of $g(t_i)$ with
$g(s_i)$ from the right combination. All this is treated in the mentioned textbooks and goes beyond the
scope of the present modest account of vector fields.



\underline{Textbooks}. Rigorous proofs lie beyond the scope of this paper. They can be found 

\begin{itemize}
\item[a)] for
mathematicians with a view on physics, for instance in Y.~Choquet-Bruhet et al.  [T.1],
especially Chapter~III,  Sections~A and B. Sections C and D offer a generalization to
several-parameter Lie groups with vector fields $V_i, i=1 \ldots n$, as infinitesimal generators, n =
dim Lie Algebra = dim G. For $n=\infty$, see Chapter~VII, Section~A.

\item[b)] for mathematicians, in several treatises including this subject: for example K.~Yano 
[T.2] Chapters~I to VII included; S.~Helgason [T.3], Chapters~I and II (there 
 our $V$ is denoted by $X$, and our $\sigma_t$ by $\gamma (t)$). Propositions 5.3 and theorem 6.1 of
Chapter~I are cornerstones to the rigorous proofs. Chapter~II offers the generalization from
one-parameter to several-parameter Lie group algebras (\S 1). See also S.~Kobayashi and T.~Nomizu
[T.4].

\item[c)] For physicists, in oversimplified compendiums of differential geometry, like, for instance
some chapters of [T.5], with definitions of the concepts used here and some sketches of proofs.
A valuable reading is also A.~Visconti [T.6].
\end{itemize}
\begin{itemize}

\item[{[T.1]}]  {Y.  Choquet-Bruhat, C. De Witt-Morette and M. Dillard-Bleick,}
 {Analysis, Manifolds and Physics}, (North-Holland, Amsterdam, 1977 and subsequent editions, 1982
etc.).

\item[{[T.2]}]  {K. Yano,}
 {The theory of Lie derivatives and its applications}, (North-Holland, Amsterdam, 1957).

\item[{[T.3]}]  {S. Helgason},
 {Differential geometry, Lie groups and Symmetric spaces}, (Academic Press, New  York, 1978).

\item[{[T.4]}]  {S. Kobayashi and K. Nomizu,}
 {Foundations of Differential Geometry, I, II} (Wiley Interscience, New York, 1963).

\item[{[T.5]}] {R. Bertlmann,}
 {Anomalies in Quantum Field Theory} (Clarendon Press, Oxford, 1996).

\item[{[T.6]}] {A. Visconti,}
 {Introductory differential geometry for physicists} (World Scientific, Singapore, 1992).
\end{itemize}

\end{document}